\documentclass[amsmath,amssymb,aps,prl,twocolumn,showpacs,superscriptaddress,groupedaddress,nofootinbib,preprintnumbers]{revtex4}

\usepackage{graphicx}
\usepackage{bm}

\newcommand{\dd}{\mathrm{d}}
\newcommand{\ee}{\mathrm{e}}

\newcommand{\del}{\partial}

\begin{document}


\author{Chul-Moon Yoo}\email{yoo@gravity.phys.nagoya-u.ac.jp}
\affiliation{
Gravity and Particle Cosmology Group,
Division of Particle and Astrophysical Science,
Graduate School of Science, Nagoya University, 
Nagoya 464-8602, Japan
}
\affiliation{
Yukawa Institute for Theoretical Physics, Kyoto University, 
Kyoto 606-8502, Japan
}

\author{Hirotada Okawa}\email{hirotada.okawa@ist.utl.pt}
\affiliation{ 
CENTRA, Departamento de F\'isica, Instituto Superior T\'ecnico, 
Av. Rovisco Pais no 1, 1049-001, Lisboa Portugal
}

\author{Ken-ichi Nakao}\email{knakao@sci.osaka-cu.ac.jp}
\affiliation{DAMTP, Centre for Mathematical Sciences, University of Cambridge, Wilberforce
Road, Cambridge CB3 0WA, United Kingdom
}
\affiliation{ 
Department of Mathematics and Physics,
Graduate School of Science, Osaka City University,
3-3-138 Sugimoto, Sumiyoshi, Osaka 558-8585, Japan
}

\vskip2cm
\title{Black Hole Universe \\ {\it -- Time Evolution --}}

\begin{abstract}
Time evolution of a black hole lattice universe 
is simulated. 
The vacuum Einstein equations in 
a cubic box with a black hole at the origin 
are numerically solved with periodic boundary conditions 
on all pairs of 
faces opposite to each other. 
Defining effective scale factors by using the 
area of a surface and the length of an edge 
of the cubic box,
we compare them with that in the Einstein-de Sitter universe. 
It is found that the behaviour of the effective scale factors is 
well approximated by that in the Einstein-de Sitter universe. 
In our model, if the box size is sufficiently larger than the 
horizon radius, local inhomogeneities 
do not significantly affect the 
global expansion law of the universe 
even though the inhomogeneity is extremely nonlinear. 
\end{abstract}

\preprint{OCU-PHYS:382}
\preprint{AP-GR:105}
\preprint{YITP-13-45}


\maketitle
\pagebreak



The effect of local inhomogeneity on the global expansion of the universe 
has drawn much attention as one of fundamental issues in 
relativistic cosmology for many years. 
One remarkable work has been done by 
Lindquist and Wheeler in 1957\cite{RevModPhys.29.432}. 
%
In this work, they investigated the so-called ``black hole lattice universe" 
composed of $N$ cells($N=$5, 8, 16, 24, 120 and 600) 
on the three sphere each of which has a black hole at the center. 
Although the lattice universe is not an exact solution, 
with help from an intuitive idea, 
it is shown that the radius of the three 
dimensional 
sphere at maximum expansion asymptotes to that of the homogeneous and 
isotropic 
dust dominated closed 
universe for a large value of $N$. 

Recently, similar inhomogeneous universe models 
to the black hole lattice universe have been 
investigated by using analytical or numerical 
techniques\cite{Clifton:2009jw,Clifton:2012qh,Uzan:2010nw,
Bentivegna:2012ei,Bruneton:2012cg,Yoo:2012jz,Bentivegna:2013xna}.
In this paper, we 
numerically construct an expanding inhomogeneous universe 
model which is composed of regularly aligned 
black holes with an identical mass $M$, 
and compare the cosmic expansion rate of this inhomogeneous universe model 
with that of the homogeneous and isotropic universe. 
Hereafter, we call this inhomogeneous 
universe model 
``black hole universe". 

In this paper, we use the geometrized units 
in which the speed of light and Newton's gravitational constant 
are one, respectively. 
The Greek indices represent spacetime components, 
whereas the small Latin indices represent spatial components. 


The way 
to construct 
the initial data 
of black hole universe 
is described in Ref.~\cite{Yoo:2012jz}. 
We briefly summarise it. 
We adopt the Cartesian coordinate system 
$\bm{x}=(x^1,x^2,x^3)$
and focus on a cubic region 
$-L\leq x^i\leq L$ ($i=1,2,3$) 
with a non-rotating black hole at the origin. 
We call the faces of this cubic box the {\it boundary} in this paper. 
Because of the discrete symmetry, 
using conventional decomposition of the Einstein equations, 
we can reduce the constraint equations to three coupled Poisson equations 
with reflection boundary condition in the region 
$0\leq x^i\leq L$. 
One of the equations is the Hamiltonian constraint equation 
for the conformal factor and the others are momentum constraint 
equations for longitudinal parts of the extrinsic curvature. 
To solve the equations, we need to fix the functional form of the 
trace part of the extrinsic curvature $K$. 
By the volume integral of 
the Hamiltonian constraint equation, we 
get an integrability condition which 
implies that there must be a domain with non-vanishing $K$. 
Since it is the simplest and the most  
convenient for our purpose to select $K$ 
vanishing in the 
neighbourhood 
of the origin, 
we adopt the following functional form of $K$: 
\begin{equation}
K(\bm x)=-3H_{\rm eff}W(R), 
\end{equation}
where $H_{\rm eff}$ is a positive constant 
which corresponds to the effective Hubble parameter,
$R:=|\bm x|$, and 
\begin{equation}
W(R)=
\left\{
\begin{array}{ll}
0&{\rm for}~0\leq R \leq \ell \\
\sigma^{-36}[(R-\sigma-\ell)^6-\sigma^6]^6&{\rm for}~\ell\leq R \leq \ell 
+ \sigma\\
1&{\rm for}~\ell+\sigma \leq R \\
\end{array}\right.,
\end{equation}
with $\ell$ and $\sigma$ being constants 
which satisfy $\ell<\ell+\sigma<L$. 
An appropriate extraction of $1/R$ divergence of 
the conformal factor allows us to solve the 
coupled Poisson equations numerically. 
It should be noted that the value of $H_{\rm eff}$ 
must be 
determined so that the 
integrability condition is satisfied, that is, 
we cannot freely choose the parameter $H_{\rm eff}$ 
but must appropriately fix the value during the numerical iteration. 
Since the results are not significantly dependent on 
$\ell$ and $\sigma$,\footnote{We set $\ell=0.1M$ and $\sigma=1.8M$ in our calculation. } 
the physical dimensionless parameter which characterises the initial data
is only $L/M$. 
In Ref.~\cite{Yoo:2012jz}, we could get convergence for $1.4M\leq L \leq 10M$. 
We adopt the case $L=2M$ as the initial data for the evolution in this paper.


The numerical time evolution is 
followed  
by COSMOS code which 
is an Einstein equation solver written in C++ by means of the 
BSSN formalism~\cite{Shibata:1995we,Baumgarte:1998te}. 
The algorithm is based on SACRA code~\cite{Yamamoto:2008js}. 
The 4th order finite differencing in space with uni-grid
and 4th order time integration with a Runge-Kutta method 
in Cartesian coordinates are adopted in COSMOS code. 
An apparent horizon solver based on Ref.~\cite{Shibata:1997nc} 
is implemented. 

We write the line element of the spacetime as 
\begin{equation}
\dd s^2=-N^2\dd t^2+\gamma_{ij}\left(\dd x^i+\beta^i\dd t\right)
\left(\dd x^j+\beta^j\dd t\right). 
\end{equation}
In order to determine the time 
slicing, 
we adopt the following condition:
\begin{equation}
\left(\frac{\del}{\del t}-\beta^i\frac{\del}{\del x^i}\right)N=-2N\left(K-K_{\rm c}\right), 
\end{equation}
where $K_{\rm c}$ is the trace of 
the extrinsic curvature at the vertex 
$x^i=L~(i=1,2,3)$ 
at each time step. 
As for the spatial coordinates, we adopt 
the so-called hyperbolic gamma driver~\cite{Alcubierre:2002kk} with specific values of parameters 
given by 
\begin{eqnarray}
\frac{\del\beta^i}{\del t}&=&B^i,\\
\frac{\del B^i}{\del t}&=&\frac{\del\tilde \Gamma^i}{\del t}-\frac{3}{4M}B^i,
\end{eqnarray}
where $\tilde \Gamma^i:=-\del_j\tilde \gamma^{ij}$ with 
$\tilde \gamma^{ij}:=\ee^{-4\psi}\gamma^{ij}$ and $\psi:=\frac{1}{12}\ln(\det\gamma)$. 

We define the cosmic expansion rate 
by using boundary variables on the geodesic 
slices~\cite{Smarr-York} which is occasionally called constant proper time slices. 
In order to know the geometry of the boundary on the geodesic slices,  
we need to solve timelike geodesic equations on the boundary. 
3+1 decomposition of geodesic equations 
is clearly described in Ref.~\cite{Vincent:2012kn}. 
Using the unit vector $n_\mu:=-N\del_\mu t$ normal to the timeslices, 
we can decompose 
the unit tangent vector field of the timelike geodesic congruence as follows: 
\begin{equation}
u^\mu=E\left(n^\mu+V^\mu\right),
\end{equation}
where $V^\mu n_\mu=0$. 
Let us represent a timelike geodesic by $x^\mu=x^\mu(t)$, where $t=x^0$. 
Then, we have~\cite{Vincent:2012kn}
\begin{eqnarray}
\frac{\dd E}{\dd t}&=&EV^i\left(NK_{ij}V^j-\del_iN\right),
\label{geoeq1}\\
\frac{\dd V^i}{\dd t}&=&NV^j\left[V^i\left(\del_j \ln N-K_{jk}V^k\right)\right.\cr
&&\left.+2K^i_{~j}-V^k \Gamma^i_{jk}\right]-\gamma^{ij}\del_jN-V^j\del_j\beta^i,
\label{geoeq2}
\end{eqnarray}
where $\Gamma^i_{jk}$ is the Christoffel symbol for the spatial metric $\gamma_{ij}$,   
and 
\begin{eqnarray}
\frac{\dd}{\dd t}&:=&\frac{N}{E}u^\mu\frac{\partial}{\partial x^\mu}
=N\left(n^\mu\frac{\del}{\del x^\mu}+V^i\frac{\del}{\del x^i}\right)\cr
&=&
\frac{\del}{\del t}+\left(-\beta^i+NV^i\right)\frac{\del}{\del x^i}. 
\end{eqnarray}
The relation between the proper time $\tau$ and the time coordinate is 
given by 
\begin{equation}
\frac{\dd \tau}{\dd t}=1/u^0=\frac{N}{E}. 
\label{propt}
\end{equation}
The initial conditions for geodesic equations are 
given by $E=1$, $V^i=0$ and $\tau=0$ in our simulation. 
It is sufficient for our purpose to consider the geodesics with 
$x^3=L$ at $\tau=0$. By the symmetry of this system, $x^3=L$ and $V^3=0$ 
always hold. 
Solving Eqs.~\eqref{geoeq1}, \eqref{geoeq2} and \eqref{propt}, 
we 
obtain 
the proper time 
in the form of 
$\tau=f(t,x^A)$ ($A=1,2$). 
Then, the line element $\dd l^2$ of the boundary $x^3=L$ 
on a hypersurface with a constant $\tau$, 
i.e., a geodesic slice 
is given by 
\begin{eqnarray}
\dd l^2&=&\Biggl[\left(-N^2+\beta^C\beta_C\right)\frac{\del_A f\del_B f}{\left(\del_t f\right)^2} \cr
&&\hspace{1.5cm}+\gamma_{AB}
-\frac{2\beta_A\del_Bf}{\del_t f}\Biggr]\dd x^A\dd x^B. 
\end{eqnarray}
We numerically obtain 
the proper area $\mathcal A$ 
of the surface 
$(-L\leq x^A\leq L,x^3=L)$ 
and the proper length $\mathcal L$ of the edge 
$(-L\leq x^1\leq L, x^2=L,x^3=L)$ 
by using this induced metric.  
Then, we define the effective scale factors as functions of the proper time as follows:
\begin{equation}
a_{\mathcal L}(\tau):=\mathcal L(\tau)~,~a_{\mathcal A}(\tau):=\sqrt{\mathcal A(\tau)}. 
\label{a-def}
\end{equation}

Since the coarse-grained black hole universe can be regarded as a 
homogeneous and isotropic universe model with vanishing spatial curvature 
by its construction, we compare the expansion rate of the black hole universe 
with that in the Einstein-de Sitter(EdS) universe.  
To compare 
the effective scale factors defined by Eq.~(\ref{a-def}) 
with that in the EdS universe, 
we define a fiducial scale factor. 
The general form of the scale factor in the EdS universe is given by 
\begin{equation}
a_{\rm EdS}(\tau):=
a_{\rm f}\left(\tau+\tau_{\rm f}\right)^{2/3}. 
\end{equation}
Then we 
determine
two parameters $a_{\rm f}$ and $\tau_{\rm f}$  
by 
the least-square fitting of this function with the numerical data in a reliable region of $\tau$. 
Using this expression, the effective Hubble parameter is given by 
\begin{equation}
H_{\rm EdS}(\tau)=\frac{2}{3}\frac{1}{\tau+\tau_{\rm f}}. 
\end{equation}


We performed the numerical simulation with four different resolutions:{\tt R1}, 
{\tt R2}, {\tt R3} and {\tt R4}. 
The interval of coordinate grids for each numerical run 
is given by 
$\Delta x/M\simeq 4/59$, $4/91$, $4/123$ and $4/155$ for {\tt R1}, {\tt R2}, {\tt R3} 
and {\tt R4}, respectively. 

\begin{figure}[htbp]
\begin{center}
\includegraphics[scale=1]{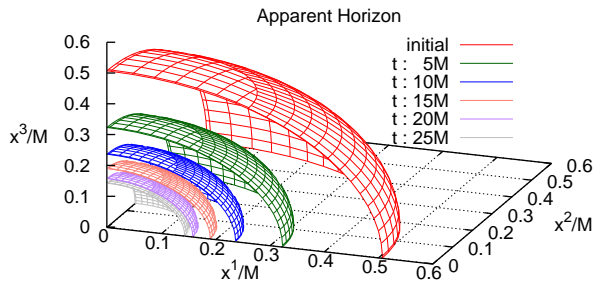}
\caption{Apparent horizon for each time step. 
}
\label{fig:ahfig}
\end{center}
\end{figure}

Before showing the results of effective scale factors, 
we note the resolution of the apparent horizon and constraint violation. 
At the initial moment, 
an almost spherical apparent horizon 
with the coordinate radius $R\simeq0.5M$ 
exists. 
As shown in Fig.~\ref{fig:ahfig}, the coordinate radius of the 
apparent horizon decreases 
as it evolves. 

At some time, our apparent horizon finder 
fails to find out the horizon because of 
the limitation of the resolution. 
For {\tt R4}, it happens at $t\sim25M$. 
While the horizon can be found, 
variation of the horizon area is smaller than 0.1\%, 
which is the same order as the typical fraction of violation of the 
Hamiltonian constraint as is shown below. 
This implies that 
the horizon area is constant in time within our numerical precision. 

\begin{figure}[htbp]
\begin{center}
\includegraphics[scale=1.2]{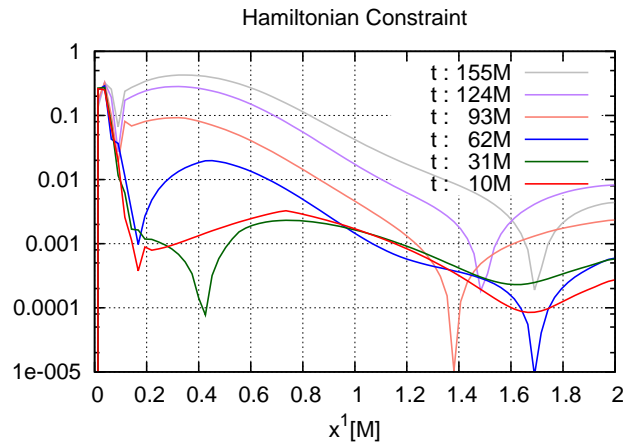}
\caption{Evolution of 
the violation of the Hamiltonian constraint  
on the $x^1$-axis for {\tt R4}. Zero means no violation 
whereas one corresponds to the largest violation.  
}
\label{fig:hamicon}
\end{center}
\end{figure}

\begin{figure}[htbp]
\begin{center}
\includegraphics[scale=1]{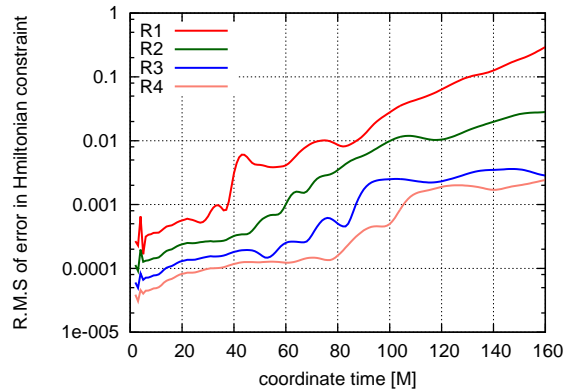}
\caption{Root-mean-square of the errors in the Hamiltonian constraint at 
grid points on the boundary. 
}
\label{hamconboundary}
\end{center}
\end{figure}

As is shown in Fig.~\ref{fig:hamicon}, 
after the failure in the apparent horizon search, 
the violation of the Hamiltonian constraint propagates 
outward, and the reliable region becomes narrower 
as time goes on. 
However, the numerical computation does not 
crash and we can proceed with the calculation. 
Even after the size of apparent horizon becomes too small to be 
resolved, 
the spacetime in the vicinity of the boundary 
is simulated with sufficient accuracy, 
since the present numerical results on the 
behaviours 
of the effective scale factors pass the convergence test. 
In order to show the convergence clearly, 
we depict the root-mean-square value of errors in the 
Hamiltonian constraint at grid points on the boundary in Fig.~\ref{hamconboundary}. 
It 
is seen from this figure 
that the error 
reduces with higher spatial resolution. 
In the beginning of the simulation, we 
find accurate second order convergence 
that the error scales to $\Delta x^2$. 
This is because the initial data sets are given by 
the 2nd order Successive-Over-Relaxation method 
for each resolution\cite{Yoo:2012jz}, 
although the evolution code has the 4th order accuracy. 
%
 
\begin{figure}[htbp]
\begin{center}
\includegraphics[scale=1]{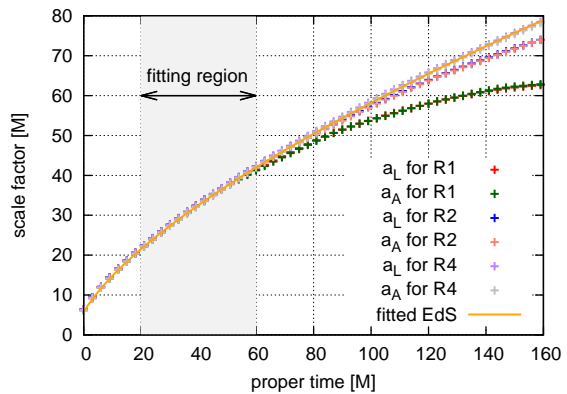}
\caption{Effective scale factors and $a_{\rm EdS}$ as functions of $\tau$. 
$a_{\mathcal A}$ almost overlaps $a_{\mathcal L}$. 
}
\label{fig:prop}
\end{center}
\end{figure}

\begin{figure}[htbp]
\begin{center}
\includegraphics[scale=1]{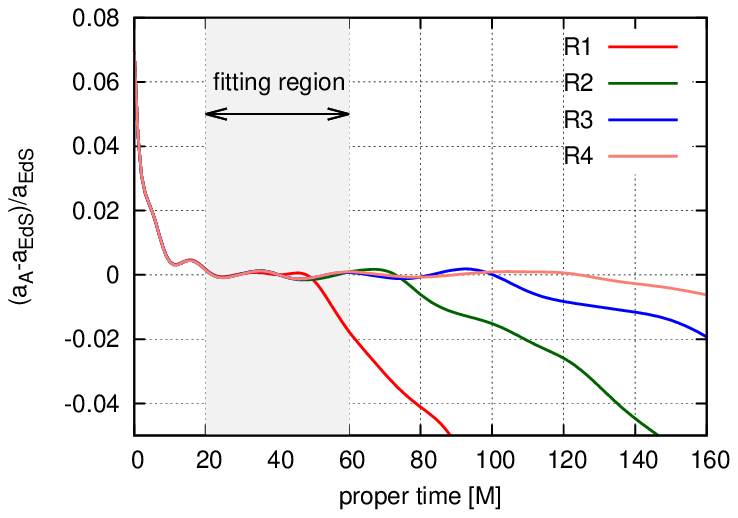}
\includegraphics[scale=1]{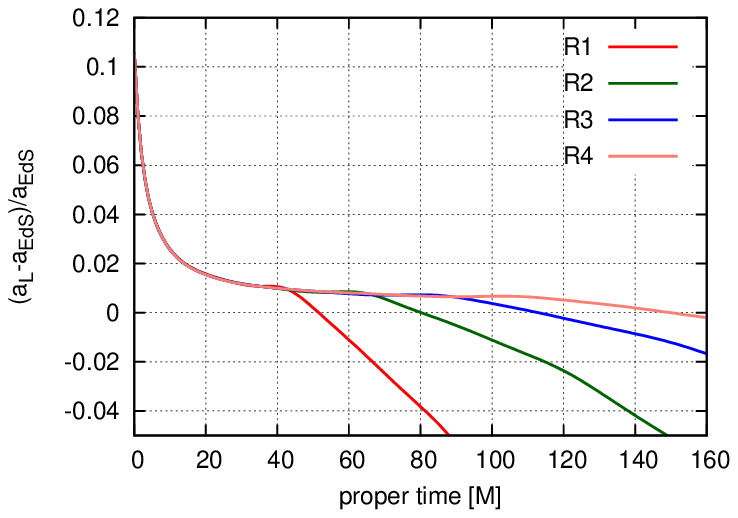}
\caption{Deviation of the effective scale factors $a_{\rm A}$ and $a_{\rm L}$ 
from $a_{\rm EdS}$ for each resolution.
}
\label{propdev}
\end{center}
\end{figure}

In Fig.~\ref{fig:prop}, 
we show the effective scale factors as functions of 
the proper time. 
The fiducial scale factor $a_{\rm EdS}(\tau)$ determined by fitting with {\tt R4} is also shown. 
The fitting is done in the region $20M<\tau<60M$, 
in which 
the results of 
all runs almost coincide with each other. 
The given value of parameters are $a_{\rm f}\simeq2.64 M$ 
and $\tau_{\rm f}\simeq-3.25 M$. 
We also show the deviation of the effective scale factors 
from 
$a_{\rm EdS}(\tau)$ in Fig.~\ref{propdev}. 
%
%
It can be found 
from Fig.~\ref{fig:prop} and \ref{propdev} that the behaviour of the 
effective scale factors for {\tt R4} is well approximated by 
that of the EdS universe in the shown period of $\tau$ 
except for a short period in the beginning. 
This result is consistent with the suggestion given by 
the numerical simulation of the collapsing eight-black-hole 
lattice\cite{Bentivegna:2012ei} and 
an initial data sequence\cite{Yoo:2012jz}. 

It would be worth 
to note the
reason why we 
have defined the effective scale factor by Eq.~\eqref{a-def}. 
The reason is that, since the total volume inside the box is infinite, 
it is hard to define the volume average in a physically 
meaningful manner. 
Therefore we do not use the spatial volume average to define the effective scale factor 
differently from other many related works(see e.g. Ref.~\cite{Rasanen:2011ki} and references therein). 

The number of black holes $N_{\rm BH}$
inside the Hubble radius $H_{\rm EdS}^{-1}$ is given by
\begin{equation}
N_{\rm BH}\sim \frac{1}{M} \times \frac{4\pi}{3}H_{\rm EdS}^{-3}
\times \frac{3}{8\pi}H_{\rm EdS}^2
=\frac{H_{\rm EdS}^{-1}}{2M}
=\frac{3}{4}\frac{\tau+\tau_{\rm f}}{M},
\end{equation}
and we obtain $N_{\rm BH}\sim 100$ for $\tau\sim 130M$. 
As was discussed in Ref.~\cite{Yoo:2012jz}, 
it is expected that the deviation of the black hole universe 
from the EdS one becomes the smaller, the larger $N_{\rm BH}$ becomes,  
or in other words, the effects of local inhomogeneities on the global expansion law 
is negligible if $N_{\rm BH}$ is sufficiently large. 
If so, the expansion law of the black hole universe will approach to that of 
the EdS universe as it expands, and the present result supports this expectation.  

The universe model studied here is highly idealised, since this work is the first step 
in the study of the universe dominated by black holes. 
The generalisations of our toy model to that with varieties of mass and 
separation etc. and the dependence of the results on them are 
still open issues which should be clarified in the future.

\section*{Acknowledgements}
We thank M.~Sasaki for encouraging us to 
work on this subject, and M.~Shibata  
for helpful discussions and comments. 
We also thank T.~Hiramatsu for 
technical supports in numerical computations. 
The numerical calculations were partly 
carried out on SR16000 at  YITP in Kyoto University.
This work was supported in part by JSPS Grant-in-Aid for Scientific
Research (C) (No. 21540276 and No. 25400265).


\end{document}